# Spatially resolved femtosecond pump-probe study of topological insulator Bi$_2$Se$_3$

Nardeep Kumar,[1] Brian A. Ruzicka,[1] N. P. Butch,[2] P. Syers,[2] K. Kirshenbaum,[2] J. Paglione,[2] and Hui Zhao[1]*

[1]*Department of Physics and Astronomy, The University of Kansas, Lawrence, Kansas 66045, USA*
[2]*Center for Nanophysics and Advanced Materials, Department of Physics,*
*University of Maryland, College Park, Maryland 20742, USA*
(Dated: April 2, 2011)

Carrier and phonon dynamics in Bi$_2$Se$_3$ crystals are studied by a spatially resolved ultrafast pump-probe technique. Pronounced oscillations in differential reflection are observed with two distinct frequencies, and are attributed to coherent optical and acoustic phonons, respectively. The rising time of the signal indicates that the thermalization and energy relaxation of hot carriers are both sub-ps in this material. We found that the thermalization and relaxation time decreases with the carrier density. The expansion of the differential reflection profile allows us to estimate an ambipolar carrier diffusion coefficient on the order of 500 cm$^2$/s. A long-term slow expansion of the profile shows a thermal diffusion coefficient of 1.2 cm$^2$/s.

## I. INTRODUCTION

As a new novel insulating state of matter, the topological insulator behaves like a bulk insulator with a large bandgap in its interior, but with a gapless surface that is protected by time-reversal symmetry.[1–4] Two-dimensional topologically protected edge states, or quantum spin Hall effect, was predicted in 2005,[5] and soon thereafter demonstrated in HgTe quantum wells.[6] More recently, three-dimensional bulk topological insulators were also predicted[7,8] and demonstrated in several binary bismuth compounds.[9–11]

Optical techniques are standard tools used to study electronic and lattice properties of solids. Several theoretical works have illustrated interesting interactions between topological insulators and light.[12–15] Experimentally, reflection,[16,17] transmission,[17] Kerr[18] and Faraday rotations,[19] and second-harmonic generation[20] of Bi$_2$Se$_3$ have been studied in visible,[20] infrared,[17,19] and terahertz ranges.[18] Different from these steady-state measurements, time-resolved optical studies can reveal carrier and phonon dynamics in topological insulators. Very recently, an ultrafast pump-probe experiment was reported, in which oscillatory differential reflection signals are observed.[21]

Previously, ultrafast pump-probe techniques have been used to study phonon dynamics in many different solids. Generally, phonons are generated by an ultrafast pump pulse via excitation of hot carriers that rapidly relax their energy by phonon emission. These phonons can induce differential reflection or transmission signals that show oscillatory behaviors as a function of the time delay between the probe and the pump pulses. However, the oscillation can be induced by three very different mechanisms: interference between multiple probe reflections, reflections of a strain wave at sample boundaries, and coherent lattice vibrations.

If the sample is transparent at the probe wavelength, the probe pulse penetrates into the sample. If an acoustic wave is generated at the sample surface by the pump pulse and propagates into the sample, the part of the probe that is reflected by the sample surface can interfere with that reflected by the propagating acoustic wave. Hence, the overall reflection coefficient oscillates with the probe delay because the phase of the second reflection depends on the location of the propagating acoustic wave. The oscillation period is proportional to the probe wavelength and is inversely proportional to the index of refraction and the sound speed.[22,23] This technique has been used to study the propagation of acoustic waves in many material systems including As$_2$Te$_3$,[22,23] GaAs-based multilayer structures,[24–28] GaN[29] and its heterostructures,[30] SiO$_2$,[31] Sapphire,[32] NdNiO$_3$,[33] and several magnetic materials.[34–37]

For strongly absorptive materials, the probe pulse will mainly sense the regions near the surface. An oscillatory differential reflection signal can still be observed, with two different mechanisms. For thin film or multilayer structures, the acoustic wave generated at the surface propagates into the film. Multiple reflections from the back surface or interface gives rise to periodic modulations of the dielectric function near the surface, causing oscillatory differential reflection signals. In this case, the period is proportional to the layer thickness and inversely proportional to the sound speed. It is independent of the probe wavelength. This technique can be used to determine the thickness, the sound speed, and the damping of the acoustic waves in several materials including As$_2$Te$_3$,[22,23] gold[38] and its nanostructures,[39] silicon,[40,41] germanium,[40] GaAs/AlGaAs heterostructures,[27] and InAs quantum dots.[42]

Bulk materials with strong absorption do not have the previous two mechanisms. However, coherent lattice vibration caused by phonons instantaneously generated at the surface can modulate the bond length, and induce oscillatory differential reflection signals. Such oscillations come directly from lattice vibration, and therefore have a frequency solely determined by the modes of the phonons excited. Coherent phonons in many systems have been studied by this process, including GaAs/AlAs superlattices,[43–47] quantum dots of InAs,[48] PbS,[49] CdTe,[50] GaN[51] and its quantum wells,[52] cuprate thin films,[53] Bismuth nanowires,[54] silicon membranes,[55] and germanium[56]

Here we report a spatially resolved ultrafast pump-probe study on $Bi_2Se_3$ crystals. We observed similar oscillatory differential reflection signals as reported in Ref. 21 that are caused by coherent optical and acoustic phonons. In addition, the rising time of the signal shows that the thermalization and energy relaxation of hot carriers are sub-ps in this material. We found that the thermalization and relaxation time decreases with the carrier density. The spatial expansion of the differential reflection profile allows us to estimate an ambipolar carrier diffusion coefficient on the order of 500 $cm^2$/s. The long-term slow expansion shows a thermal diffusion coefficient of 1.2 $cm^2$/s.

The paper is organized as follows. In Sec. II, we describe the spatially resolved pump-probe technique and the samples used for this study. In Sec. III, the results of the measurements are presented. Discussions on the hot carrier dynamics, coherent optical phonons and coherent acoustic phonons are given in Sec. IV. In Sec. V we briefly summarize the results.

## II. EXPERIMENTAL TECHNIQUE AND PROCEDURE

The $Bi_2Se_3$ single crystals are grown via a modified Bridgman process and slow cooled in a constant temperature gradient of approximately 14 K/cm over three weeks. The carrier density in the bulk, determined from Shubnikov de Haas oscillations, is about $1.2 \times 10^{19}$ /$cm^3$. The bulk transport properties are similar to those previously observed in metallic samples with comparable carrier density, i.e., the samples are rather metallic.[57] The sample surface had been exposed to air for about two months before the optical pump-probe measurements were taken. However, the main observations are reproducible on a fresh surface obtained by cleaving the sample with an adhesive tape.

Figure 1(a) shows schematically the experimental setup. A titanium sapphire laser (Ti:Sa) provides ultrashort pulses with a central wavelength of 800 nm and a repetition rate of 80 MHz. The majority of the output is used to pump an optical parametric oscillator (OPO) with a signal output of 1480 nm. The second harmonic of this output with a central wavelength of 740 nm is obtained by using a nonlinear optical crystal (BBO). This pump pulse is focused to a Gaussian spot of 2.4 $\mu$m full width at half maximum (FWHM) at the sample surface by using a microscope objective lens with a high numerical aperture. A small fraction of the 800-nm output of the Ti:Sa is used as the probe pulse, and is focused to the sample through the same lens to a Gaussian spot of 4.0 $\mu$m. The probe spot is scanned with respect to the pump spot on the sample surface by tilting the beam-splitter that sends the probe beam to the objective lens. The time delay between the probe and the pump pulses can be controlled by moving a reflector in the probe arm.

Since the objective lens is made of dispersive materials and is rather thick, the pump and probe pulses are expected to be broadened. To determine the actual temporal widths of the pulses, we put a thin GaAs crystal grown along (110) direction at the sample location, and detect the sum-frequency generation of the two pulses as a function of the time delay between the two pulses. The intensity cross-correlation, shown as the gray area in Fig. 3(c), has a temporal width of 250 fs. This defines the time resolution of our study. Furthermore, this process allows us to accurately determined the zero probe delay, which is defined as the time when the centers of the probe and the pump pulses overlap.

The reflected probe is collimated by the objective lens, and detected by a photodiode. The differential reflection of the probe, $\Delta R/R_0 = (R - R_0)/R_0$, is defined as the relative change in the reflection ($R$) caused by the pump pulse. The $R_0$ is the reflection without the presence of the pump pulse, and is measured to be about 0.50. To measure $\Delta R/R_0$, a mechanical chopper is used to modulate the intensity of the pump pulse with a frequency of several kHz, and a lock-in amplifier slaved to the modulation frequency is used to detect the voltage of the photodiode. A balanced detection technique is used in order to improve the signal-to-noise ratio.[58] The pump and the probe pulses are linearly polarized along perpendicular directions. We have also repeated one measurement with the two pulses polarized along the same direction, and found no difference in the signal. All of the measurements are performed with the sample at room temperature and in the air.

In this two-color pump-probe scheme, the pump beam can be prevented from reaching the detector by using color filters in addition to linear polarizers. This significantly improves the signal-to-noise ratio. Furthermore, it allows study of the signal near zero delay since the direct coherent interaction between the two pulses is greatly suppressed.

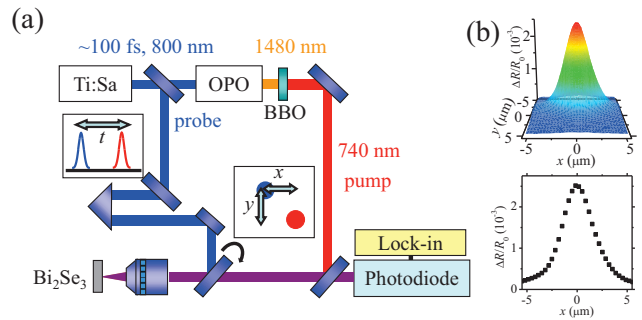

FIG. 1. Panel (a) shows schematically the experimental system used for the experiment. The upper part of Panel (b) shows the spatial profiles of the differential reflection signal measured by scanning the probe spot in the $x-y$ plane with a probe delay of 0.4 ps. The peak energy fluence of the pump pulse is 56 $\mu$J/$cm^2$. The lower part shows a cross section on the $x$ axis ($y=0$).

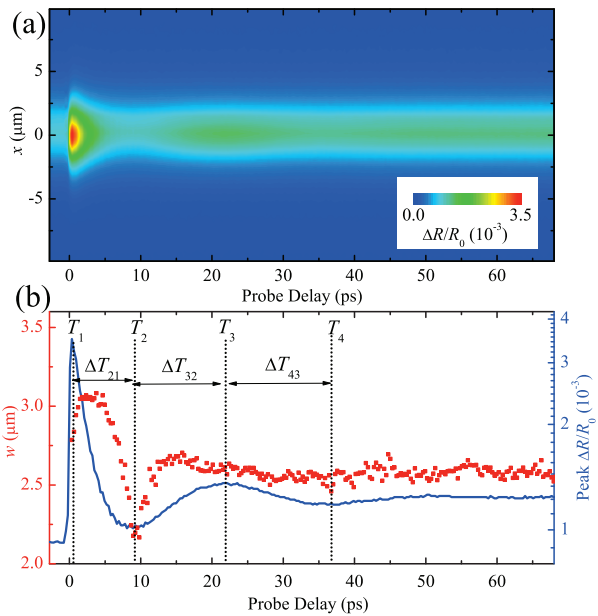

FIG. 2. (a): Differential reflection signal as a function of the probe delay and $x$, measured with $y = 0$. (b): The peak differential reflection signal ($x = y = 0$) as a function of the probe delay (solid line, right axis) showing an oscillating feature. The squares (left axis) show the width (FHWM) of the profile as a function of probe delay, obtained through a procedure shown in Fig. 3.

## III. EXPERIMENTAL RESULTS

Figure 1(b) shows the spatial distribution of the $\Delta R/R_0$ signal measured by scanning the probe spot on the sample surface with a fixed probe delay of 0.4 ps. Here the $x = y = 0$ is defined as where the centers of the pump and the probe spots overlap. The $x$ direction is arbitrarily chosen with respect to the crystal orientation, since the sample property in the $x - y$ plane is expected to be isotropic. The cross section along the $x$ axis is plotted in the lower panel of Fig. 1(b). The average power of the pump beam is 400 $\mu$W, corresponding to a peak energy fluence of 56 $\mu$J/cm$^2$ at the center of the pump spot. Using the reflection coefficient of 0.50 and assuming all of the un-reflected pump photons are absorbed near the surface, a fluence of 1 $\mu$J/cm$^2$ corresponds to an areal carrier density of $2 \times 10^{12}$/cm$^2$.

Since the spatial profile shown in Fig. 1(b) is isotropic, in the rest of the study we only measure the differential reflection signal along the $x$-axis. This allows us to acquire the $x$-profiles at many probe delays in order for a more quantitative study. Figure 2(a) shows how the differential reflection signal varies with the probe delay and $x$, with a fixed $y = 0$. The peak energy fluence of the pump pulse for this measurement is 78 $\mu$J/cm$^2$. In the whole time range, the shape of the profile remains the same. The solid line in Fig. 2(b) (right axis) shows a cross section of Fig. 2(a) with $x = 0$. Pronounced os-cillation of the differential reflection signal is observed. For later discussions, we define some characteristic times in Fig. 2(b), where $T_i$ is the time for the $i$-th extreme value, and $\Delta T_{ji}$ the time difference between the $i$-th and $j$-th extremes. The differential reflection signal reaches a peak at $T_1$ that is slightly after zero delay (0.4 ps). After a period of $\Delta T_{21} = 8$ ps, the signal reaches a minimum at $T_2$. After that, the signal increases for a period of $\Delta T_{32} = 14$ ps followed by another decrease for a period of $\Delta T_{43} = 15$ ps. After $T_4 = 37$ ps, the signal increases again, until about 50 ps, when the signal reaches a steady level. The fact that $\Delta T_{43} > \Delta T_{32} > \Delta T_{21}$ indicates that the signal oscillates with a decreasing frequency. The period of $\Delta T_{21} + \Delta T_{32} = 22$ ps corresponds a frequency of 0.045 THz and an energy of 0.19 meV.

We note in Fig. 2 that a nonzero differential reflection signal is detected at negative probe delays, i.e. when the probe pulse arrives earlier than the pump pulse. This shows that the sample does not fully recover between the pump pulses, which arrive in an interval of about 12.5 ns. This signal is caused by the previous pump pulse. Hence, the differential reflection signal measured with a positive probe delay is composed of the signal caused by the present pump pulse and such a background. Since the background signal is relatively steady, we can subtract it from the measured signal to obtain the signal caused by the present pump pulse. Figure 3(a) shows such a procedure. The highest curve is measured with a positive probe delay of 0.35 ps, while the lowest curve is obtained by averaging 20 profiles measured at negative probe delays (several ps before zero delay). By subtracting the latter from the former, we obtain the differential reflection signal associated with the present pump pulse, which is plotted as the middle curve in Fig. 3(a).

In our experiment, we independently determined that the probe spot has a Gaussian shape with a width of 4 $\mu$m. The measured profiles are results of convolutions of the probe spot and the actual profile of the differential reflection. Therefore, we fit all the profiles with a Voigt function with a fixed Gaussian width of 4 $\mu$m. The width of the actual profiles (Lorentz) is plotted in Fig. 2(b) as the red squares (left axis). Pronounced oscillations in the width are observed, with some extremes occurring at the same time with the extremes in the height (blue solid line). In order to verify that the oscillation in width is not an artifact caused by the fit process, and merely reflects the oscillation in height, we have also tried to fit the profiles with Gaussian and Lorentzian functions. These fits have larger mean-square errors, but the oscillations remains the same. As another verification, we plot four profiles (normalized) in Fig. 3(b) that are chosen in a random fashion but around the probe delays when the width is an extreme. one can directly tell that the profile broadens from 0.35 (black) to 1.78 ps (red), and then shrinks at 10.7 ps (pink), and broadens again at 15.8 ps (blue). The relative widths of these profiles determined by naked eyes are consistent with the fit shown in Fig. 2(b). The third evidence that the oscillation in width is not caused by the oscillation in height is that in the first several picoseconds, the width increases while the height drops, as clearly shown in Fig. 3(c).

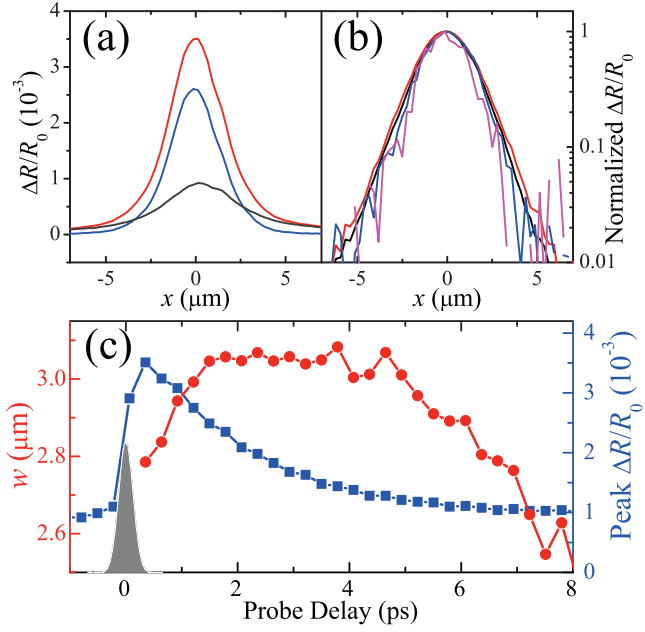

FIG. 3. Panel (a) shows the spatial profile of differential reflection with a probe delay of 0.35 ps (red/highest) and with negative delays (gray/lowest). The blue/middle curve is the difference. Panel (b) shows the profiles (after removing the background) measured at probe delays of (from wide to thin) 1.78 (red), 0.35 (black), 15.8 (blue), and 10.7 (pink), respectively. Panel (c) is the same as Fig. 2(b) but with a smaller time range near zero delay. The gray area shows the cross correlation of the pump and the probe pulses.

In order to investigate the rich dynamics of the oscillatory signal, we measure the time evolution of the height of the differential reflection signal with different pump fluences. Some examples of the results are summarized in Fig. 4(a). It can be clearly seen that the second peak ($T_3$) shifts systematically with the fluence, showing that the excitation level influences the dynamics. Quantitatively, we plot these characteristic times in Fig. 4(b) as functions of fluence. The $T_1$ (the rising time) decreases from about 0.8 to 0.3 when the fluence increases from 6 to 100 $\mu$J/cm$^2$. The interval $\Delta T_{21}$ is almost independent of the fluence. However, $\Delta T_{32}$ becomes significantly longer (from 8 to 15 ps) when increasing the fluence. Finally, $\Delta T_{43}$ is independent of the fluence, and is longer than other intervals. Figure 4(c) shows that the height of the first peak (at time $T_1$) increases linearly with the fluence.

We note that in Ref. 21 a very low pump fluence (2.5 $\mu$J/cm$^2$) was used in order to avoid heating damage of the sample. The highest pump fluence used in our experiment is 100 $\mu$J/cm$^2$. We see no sign of sample damage.

In addition to the oscillation with a frequency of about 0.045 THz shown in Figs. 2-4, a high-frequency oscillation is observed at early probe delays. Figure 5(a) shows the differential reflection signal as a function of the probe delay for the first 8 ps, with different pump fluences. Apparently, the slow oscillation of 0.045 THz is superimposed with a high-frequency oscillation. To quantitatively analyze this high-frequency component, we remove the slow varying components from the curves shown in Fig. 5(a) by using a high-pass Fourier filter with a cutoff frequency of 2.0 THz. As an example, the points in the inset of Fig. 5(a) show the high-frequency component of the top curve with a pump fluence of 50.4 $\mu$J/cm$^2$. The high-frequency components are fit by using a damped sinusoidal functions, $Ae^{-(t-t_0)/\tau}sin[2\pi f(t-t_1)]$. The result for the pump fluence of 50.4 $\mu$J/cm$^2$ is plotted as the solid line in the inset of Fig. 5(a). Figure 5(b) summarizes the initial amplitude ($A$, squares), the frequency ($f$, circles), and the decay time ($\tau$, triangles), that are obtained from the fits. The initial amplitude of the oscillation increases linearly with the pump fluence. The frequency seems to be independent of the fluence. The slight increase (less than 1%) at low fluence could be attributed to the artifact of the Fourier filter when applied to more noisy curves. By averaging the high-fluence data, we deduce a frequency of 2.167 ± 0.002 THz, which corresponds to an energy of 8.974 meV. The decay time deduced from the fits is also independent of the fluence, with an average value of 3.2 ± 0.1 ps.

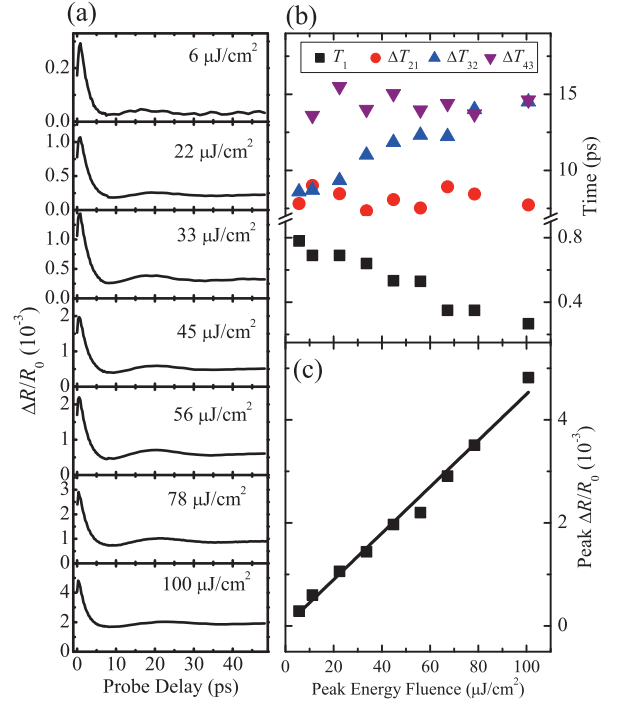

FIG. 4. Panel (a) shows the peak differential reflection signal ($x = y = 0$) as a function of the probe delay with different pump fluences as indicated in each plot. Panels (b) and (c) summarizes the characteristic times and the peak of the signal as functions of the pump fluence.

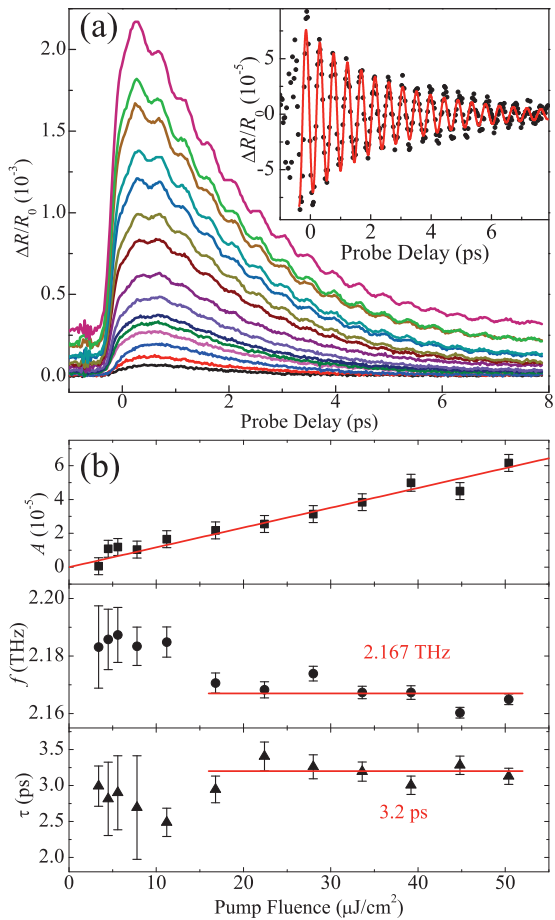

FIG. 5. Panel (a): The peak differential reflection signal ($x = y = 0$) as a function of the probe delay in the first 8 ps, with the pump fluences of (from bottom to top) 0.6, 1.1, 2.2, 3.4, 4.5, 5.6, 7.8, 11.2, 16.8, 22.4, 28, 33.6, 39.2, 44.8, and 50.4 $\mu$J/cm$^2$. The inset shows the high-frequency oscillation obtained by using a high-pass filter with a cutoff frequency of 2 THz. Panel (b) summarizes the initial amplitude (squares), the frequency (circles), and the decay time (triangles) that are obtained from the fits.

In order to examine whether these oscillation frequencies are influenced by the size of the pump laser spot, we measure the peak differential reflection signal as a function of the probe delay with two different pump spot sizes of 3 and 11 $\mu$m. The larger spot size is achieved by intentionally un-collimating the pump beam with a telescope. The pump power is adjusted so that in each case the peak fluence, i.e. the fluence at the center of the spot, is the same 17 $\mu$J/cm$^2$. As shown by the two curves in Fig. 6, similar temporal behaviors of the differential reflection are observed. The only noticeable difference is the background. Furthermore, the inset of Fig. 6 shows the two curves in early probe delays. Clearly, the high-frequency oscillation is independent of the spot size, too.

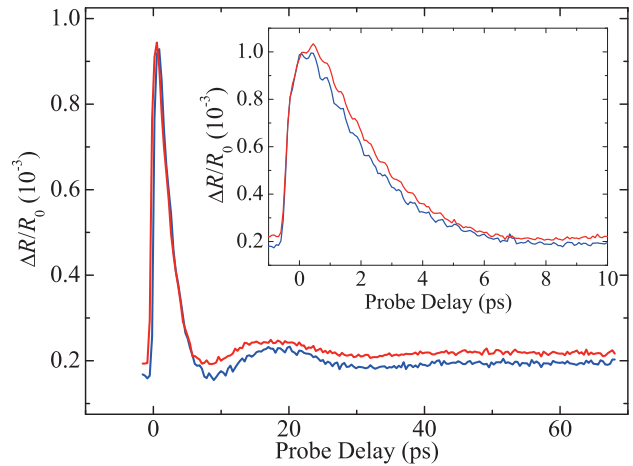

FIG. 6. The peak differential reflection signal ($x = y = 0$) as a function of the probe delay with different pump spot sizes of 3 $\mu$m (blue/lower) and 11 $\mu$m (red/upper), respectively. The energy fluence of the pump pulse at the center of the spots is 17 $\mu$J/cm$^2$ for both measurements. The inset shows the two curves at early probe delays.

## IV. DISCUSSIONS

### A. Origin of the Differential Reflection Signals

Generally, a differential reflection signal can be induced by the lattice excitations (phonons) and/or electronic excitations (carriers). Carriers can change the reflection via two major mechanisms. First, free carriers can absorb photons, causing an increase in absorption and therefore a decrease in reflection. Such an effect is usually small since the absorption is indirect, i.e., a third quasi-particle must be involved to fulfill the crystal momentum conservation. In the experiments the sign of the differential reflection is always positive. That is, the reflection is increased by the presence of the pump pulse. This is inconsistent with the free-carrier absorption. The second mechanism of carrier-induced differential reflection is the effect of phase-space filling. The carriers occupy energy states in the conduction band, reducing absorption. This mechanism gives a positive differential reflection, and can be rather strong if the probe photon energy matches the energy states occupied by the carriers. In our experiment, the probe photon energy of 1.55 eV is much higher than the bandgaps of both the bulk (0.35 eV) and the surface (0 eV). Therefore, a significant occupation of the probing states can only occur when the carriers are still hot. Similar experiments on graphene, which has a similar bandstructure with the surface states of topological insulators, have shown that the differential reflection/transmission signal only exists for at most a few ps.[59–61] Therefore, although it is possible that the signal observed at very early probe delays has a contribution from the hot carriers, at later probe delays after 10 ps, the signal is unlikely to be induced by the carriers.

As described in Sec. I, the lattice excitation can induce oscillatory differential reflection signals via three mechanisms: interference of the probe reflections by the surface and by the strain wave, multiple reflection of the strain wave, and coherent lattice vibration. Since the sample is highly absorptive with a penetration depth of less than 100 nm, the first mechanism can be safely eliminated. In the second mechanism, the oscillation frequency is solely determined by the sound speed and the thickness of the samples. Since very similar frequencies are observed in different samples[21] and the low-frequency varies with time [Fig. 4(b)], we can also safely eliminate this mechanism. The only possible mechanism is then the coherent lattice vibration.

In conclusion, we attribute the signal at probe delays on several ps or longer to the coherent lattice vibration, and the signal at early probe delays to both the coherent lattice vibration and (possibly) the hot carriers.

## B. Hot Carrier dynamics

The squares in Fig. 4(b) show a systematical variation of the rising time of the differential reflection signal as we change the pump fluence. The rising time is clearly not pulse-width limited; the time resolution of the study, given by the cross-correlation of the pump and the probe pulses [gray area in Fig. 3(c)] is shorter than all of the rising times measured. Regardless of the dominating mechanism that causes the signal on this time range, the finite rise time is caused by thermalization and energy relaxation of hot carriers. The pump photon energy is 125 meV larger than the probe. If the electrons and the holes are excited with the same excess energy, the pumping states in the conduction band are 62.5 meV higher than the probing states. If the signal in the first ps is dominated by the carrier contribution, the rise of the signal is caused by the movement of carriers from the pumping states to the probing states by thermalization and energy relaxation. The decay after the peak reflects the movement of the carriers to even lower energy states. On the other hand, if signal is dominated by the phonons, the rising time reflects the increase of phonon population. Since the phonons are emitted by the hot carriers during their energy relaxation, the rising time measures the time takes for the carriers to reach thermal equilibrium with the lattice.

Our measurement shows that the thermalization and energy relaxation of hot carriers in $Bi_2Se_3$ is ultrashort. The rising time is in the range of 0.3 to 0.8 ps. The thermalization and energy relaxation time decreases with increasing carrier density. We note that in Ref. 21 a rising time of about 1 ps is observed with a pump fluence of 2.5 $\mu J\ cm^{-2}$, which is consistent with the trend shown in Fig. 3. The density dependence can be attributed to the density-dependent optical phonon emission rate, or the density-dependent thermalization time.

The spatial expansion of the profile at early probe delays shows the diffusion of carriers. The red circles in Fig. 3(c) clearly show that the profile expands from about 2.8 $\mu m$ to more than 3.0 $\mu m$ within the first 2 ps. During this time, the height of the profile [blue squares in Fig. 3(c)] decreases. If the broadening is caused by diffusion of hot carriers or phonons. The squared width is expected to expand linearly with a slope of about $16ln(2)D$, where $D$ is the diffusion coefficient.[62] Therefore, the observed expansion corresponds to a diffusion coefficient on the order of 500 $cm^2/s$. Previous experiments have shown that the thermal diffusion coefficient of $Bi_2Se_3$ is about 0.3 $cm^2/s$,[63] three orders of magnitude smaller than what we observed. Hence, we conclude that the spatial expansion is caused by diffusion of the photoexcited hot carriers. We note that in the diffusion process the excited electrons and holes move as pairs due to the Coulomb interaction between. The quantity measured is therefore the ambipolar diffusion coefficient.

The expansion of the profile can reflect the carrier diffusion via two possible mechanisms. If the differential reflection signal is dominated by the carriers, the expansion directly measures the diffusion. If the differential reflection signal is mainly caused by phonons, the expansion is induced by phonon emission of the carriers that have diffused out of the original profile. Given the complicated origins of the signal, the agreement is reasonably good.

## C. Coherent Optical Phonons

We attribute the high-frequency oscillation in the first 10 ps shown in Fig. 5(c) to optical phonons. That is, the coherent relative vibration of atoms in each unit cell causes modulation of the reflection. The oscillation has a frequency of 2.167 ± 0.002 THz [circles in Fig. 5(b)], which corresponds to an energy of 8.974 meV. The frequency does not change with the probe delay, and is independent of the pump fluence. Such a frequency is consistent with the $A_{1g}^l$ longitudinal optical phonon frequency determined by Raman spectroscopy (2.16 THz)[64] and the previous pump-probe study (2.13 THz).[21] The linear relation between the initial amplitude of the oscillation and the pump fluence [squares in Fig. 5(b)] indicates that the excited optical phonon population is proportional to the pump fluence. The decay time of the oscillation is 3.2 ± 0.1 ps [triangles in Fig. 5(b)], and is also independent of the pump fluence. This is the lifetime of the optical phonons.

A surface plasma oscillation can be ruled out as the cause of the high-frequency oscillation since the frequency is independent of the spot size, as shown in Fig. 6, and excitation carrier density, as shown in Fig. 5(b). The frequency of the surface plasma oscillation is expected to be proportional to the square root of the carrier density divided by the width of the spot.

### D. Coherent Acoustic Phonons

We attribute the low-frequency oscillation of 0.045 THz to coherent acoustic phonons. The acoustic-mode coherent vibration of the lattice causes modulation of the reflection, with a frequency determined by the phonon energy.[43–56] The frequency corresponds to an energy of 0.19 meV. However, the frequency of the oscillation decreases significantly with time, as shown in Fig. 2(b). The $T_{43}$ (15 ps) is almost twice longer than $T_{21}$ (8 ps). Such a behavior could be attributed to energy relaxation of acoustic phonons; however, a time-varying frequency has never been observed in similar studies of other materials before.[43–56] We observed a significant increase in $T_{32}$ with the pump fluence; however, $T_{21}$ and $T_{43}$ are independent of the fluence [Fig. 4(b)].

The evolution of the spatial profile of the signal illustrates the transport properties of the acoustic phonons. The red squares in Fig. 2(b) show that, after the transient process of about 40 ps, the width reaches a quasi-steady value of 2.6 $\mu$m. The width of the background signal at negative delays is about 4.9 $\mu$m. Assuming the background is caused by the previous pump pulse that arrived about 12.5 ns earlier, the width roughly broadens from 2.6 to 4.9 $\mu$ in 12.5 ns. This gives a diffusion coefficient of acoustic phonons of 1.2 cm$^2$/s, within a factor of four with the previously reported thermal diffusion coefficient of 0.3 cm$^2$/s.[63]

The pronounced dip in the width of the profile shown in Fig. 2(b) around 8 ps, exactly when the peak differential reflection signal reaches the minimum, and the overall decrease of the width from early delays of several ps (about 3.1 $\mu$m) to later delays of about 20 ps (2.6 $\mu$m), are not understood. We speculate that it might be related to the decay of optical phonons to acoustic phonons.

## V. SUMMARY

We have shown that a spatially resolved femtosecond pump-probe technique can be used to study carrier and phonon dynamics in topological insulators. Pronounced oscillations in differential reflection on Bi$_2$Se$_3$ were observed with two distinct frequencies. The high-frequency oscillation of 2.167 THz decays in 3.2 ps, and is caused by coherent optical phonons. The frequency is independent of the probe delay, the pump fluence, or the pump spot size. The low-frequency oscillation of 0.045 THz exists for more than 50 ps, and can be attributed to coherent acoustic phonons. We found that the frequency is independent of the spot size. The rising time of the signal shows the thermalization and energy relaxation of hot carriers are sub-ps in this material, and decrease with the carrier density. The spatial expansion of the differential reflection profile allows us to estimate an ambipolar carrier diffusion coefficient on the order of 500 cm$^2$/s. The long-term slow expansion shows a thermal diffusion coefficient of 1.2 cm$^2$/s, which agrees with the previously reported value within a factor of four.

Some of the observed features in this experimental study are not fully understood yet. For example, in the low-frequency oscillation the characteristic time $T_{32}$ increases with carrier density, but not the other two characteristic times; A breathing-like movement of the profile is observed in the first 15 ps, along with an overall decrease of the profile size from several ps to about 15 ps. More experimental and theoretical works are needed to understand these features.

### ACKNOWLEDGMENTS

We acknowledge many useful discussions with H. D. Drew of University of Maryland. NK, BAR, and HZ acknowledge support from the NSF (DMR-0954486). Work at the University of Maryland was supported by NSF MRSEC (DMR-0520471) and DARPA-MTO award (N66001-09-c-2067). NPB was partially supported by CNAM.